\documentclass[useAMS,usenatbib]{mn2e}
\usepackage{amsmath}
\usepackage{txfonts}
\usepackage{graphics}
\usepackage{times}

\renewcommand{\vec}[1]{\mathbf{#1}}
\newcommand{\vr}{\vec{r}}
\newcommand{\vq}{\vec{q}}
\newcommand{\vx}{\vec{x}}
\newcommand{\vk}{\vec{k}}
\newcommand{\vPsi}{\vec{\Psi}}
\newcommand{\vDelta}{\vec{\Delta}}
\newcommand{\hq}{\hat{q}}
\newcommand{\hk}{\hat{k}}
\newcommand{\hz}{\hat{z}}
\newcommand{\mJ}{\mathcal{J}}
\newcommand{\lam}{\lambda}

\begin{document}


\title[The Zeldovich approximation]{The Zel'dovich approximation}
\author[White]{
    Martin White$^{1,2}$ \\
    $^{1}$ Departments of Physics and Astronomy, University of California,
    Berkeley, CA 94720, USA \\
    $^{2}$ Lawrence Berkeley National Laboratory, 1 Cyclotron Road,
    Berkeley, CA 94720, USA
}
\date{\today}
\pagerange{\pageref{firstpage}--\pageref{lastpage}}

\maketitle

\label{firstpage}

\begin{abstract}
This year marks the $100^{\rm th}$ anniversary of the birth of Yakov Zel'dovich.
Amongst his many legacies is the Zel'dovich approximation for the growth of
large-scale structure, which remains one of the most successful and insightful
analytic models of structure formation.  We use the Zel'dovich approximation
to compute the two-point function of the matter and biased tracers, and compare
to the results of N-body simulations and other Lagrangian perturbation theories.
We show that Lagrangian perturbation theories converge well and that the
Zel'dovich approximation provides a good fit to the N-body results
except for the quadrupole moment of the halo correlation function.
We extend the calculation of halo bias to $3^{\rm rd}$ order and also consider
non-local biasing schemes, none of which remove the discrepancy.
We argue that a part of the discrepancy owes to an incorrect prediction of
inter-halo velocity correlations.
We use the Zel'dovich approximation to compute the ingredients of the Gaussian
streaming model and show that this hybrid method provides a good fit to
clustering of halos in redshift space down to scales of tens of Mpc.
\end{abstract}



\section{Introduction}
\label{sec:intro}

This year marks the $100^{\rm th}$ anniversary of the birth of
Yakov Zel'dovich, who was a pioneer in the study of large-scale structure
and introduced the approximate dynamics that bears his name \citep{Zel70}.
The Zel'dovich approximation provides an intuitive way to understand the
emergence of the beaded filamentary structure which has become known as the
cosmic web and a fully realized (though approximate) model of non-linear
structure formation \citep{Pee80,ColLuc95,Pea99}.
The Zel'dovich approximation predicts the rich structure of voids, clusters,
sheets and filaments observed in the Universe \citep{Dor80,PauMel95},
and indeed it provides a reasonably good match to N-body simulations on
large scales \citep{CMS93,TasZal12a,TasZal12c}.
For a discussion of why the Zel'dovich approximation works so well,
see \citet{Buc89,PauMel95,YMGM,Tas14a}.
For reviews of the Zel'dovich approximation, see the textbooks referenced
above and
\citet{ShaZel89,SahCol95,ColSah96,GurSaiSha12,HidShaWey14}.

The last few years have seen a resurgence of interest in the Zel'dovich
approximation.  It has been applied to understanding the effects of non-linear
structure formation on the baryon acoustic oscillation feature in the
correlation function \citep{PadWhi09,McCSza12,TasZal12a} and to
understanding how ``reconstruction'' \citep{Eis07} removes those
non-linearities \citep{PadWhiCoh09,NohWhiPad09,TasZal12b}.
It has been used as the basis for an effective field theory of large-scale
structure \citep{PorSenZal14}.
It has been compared to ``standard'' perturbation theory \citep{Tas14a},
extended to higher orders in Lagrangian perturbation theory
\citep{Mat08a,Mat08b,OkaTarMat11,CLPT}
and to higher order statistics \citep{Tas14b}.
Despite the more than 40 years since it was introduced, the Zel'dovich
approximation still provides one of our most accurate models for the
distribution of cosmological objects.

In this paper we investigate to what extent the Zel'dovich approximation can
be used as a quantitatively accurate model of the low-order clustering of
objects in cosmology.
The outline is as follows.
After some background and review to establish notation in
Section \ref{sec:review}
we present a derivation of the 2-point function within the Zel'dovich
approximation \citep[see also][]{CLPT,Tas14a,Tas14b} both for matter
(Section \ref{sec:matter}) and for biased tracers (Section \ref{sec:biased}).
In these sections we show that the principle ingredient to the calculation,
the Lagrangian correlator, can be inverted analytically and thus the
correlation function expressed as a simple quadrature.
All of the ingredients to the Zel'dovich approximation involve only one
dimensional integrals of the linear theory power spectrum, and these can be
efficiently precomputed and tabulated, making numerical evaluation fast and
efficient.
We compare the Zel'dovich calculation to some other Lagrangian perturbation
theory schemes, and to the results of N-body simulations, in
Section \ref{sec:results}.
We extend previous calculations of the effects of bias to higher order and
include non-local, Lagrangian bias in Section \ref{sec:nonlocal}.
Finally we introduce the Zel'dovich Streaming Model (ZSM) as a hybrid
method for accurately computing the redshift-space correlation function of
biased tracers in Section \ref{sec:zsm} and end with a discussion of our
results in Section \ref{sec:discussion}.

For plots and numerical comparisons we assume a $\Lambda$CDM cosmology with
$\Omega_m = 0.274$, $\Omega_\Lambda = 0.726$, $h = 0.7$, $n = 0.95$, and
$\sigma_8 = 0.8$.  Our simulation data are derived from a suite of 20 N-body
simulations run with the TreePM code described in \citet{TreePM}.
Each simulation employed $1500^3$ equal mass
($m_p\simeq 7.6\times10^{10}\,h^{-1}M_\odot$)
particles in a periodic cube of side length $1.5\,h^{-1}$Gpc
as described in \citep{ReiWhi11,Whi11}.
Halos are found using the friends-of-friends method, with a linking length
of $0.168$ times the mean inter-particle spacing.

\section{Background and review}
\label{sec:review}

In this section we provide a brief review of cosmological perturbation theory,
focusing on the Lagrangian formulation\footnote{See \citet{Ber02} for a
comprehensive (though somewhat dated) review of Eulerian perturbation theory.}
\citep{Buc89,Mou91,Hiv95,TayHam96}.
This material should be sufficient to remind the reader of some essential
terminology, and to establish our notational conventions.
Our notation and formalism follows closely that in
\citet{Mat08a,Mat08b,CLPT,WanReiWhi13}
to which we refer the reader for further details.

In terms of the fractional density perturbation, $\delta$, we can write the
2-point correlation function,
\begin{equation}
    \xi(\vr) = \langle \delta(\vx) \delta(\vx+\vr) \rangle ,
\end{equation}
and its Fourier transform, the power spectrum $P(\vk)$, as
\begin{equation}
 \langle \delta(\vk) \delta(\vk') \rangle =
    (2\pi)^3 \delta_D(\vk+\vk') P(\vk) .
\end{equation}
Here $\delta_D$ denotes the 3-dimensional Dirac delta function, and we use the
Fourier transform convention
\begin{equation}
    F(\vx) = \int \frac{d^3k}{(2\pi)^3}~ F(\vk) e^{i\vk\cdot\vx} .
\end{equation}
Angle brackets around a cosmological field, e.g.\ $\langle F \rangle$, signify
an ensemble average of that quantity over all possible realizations of our
universe; in most cases of interest, ergodicity allows us to replace these
ensemble averages with spatial averages over a sufficiently large cosmic
volume.

In the Lagrangian approach to cosmological fluid dynamics, one traces the
trajectory of an individual fluid element through space and time.  For a fluid
element located at position $\vq$ at some initial time $t_0$, its position at
subsequent times can be written in terms of the Lagrangian displacement field
$\vPsi$,
\begin{equation}
\label{eq:LtoE}
    \vx(\vq,t) = \vq + \vPsi(\vq,t) ,
\end{equation}
where $\vPsi(\vq,t_0) = 0$.  Every element of the fluid is uniquely labeled by
its Lagrangian coordinate $\vq$ and the displacement field $\vPsi(\vq,t)$ fully
specifies the motion of the cosmological fluid.
Lagrangian Perturbation Theory (LPT) finds a perturbative solution for the
displacement field,
\begin{equation}
\label{eq:psin}
\vPsi(\vq,t) = \vPsi^{(1)}(\vq,t) + \vPsi^{(2)}(\vq,t)
             + \vPsi^{(3)}(\vq,t) + \cdots .
\end{equation}
The first order solution is the Zel'dovich approximation \citep{Zel70},
which shall be the focus of this paper.  Henceforth we shall denote
$\vPsi^{(1)}$ simply as $\vPsi$:
\begin{equation}
    \vPsi(\vq) = \int \frac{d^3k}{(2\pi)^3}
    \ e^{i\vk\cdot\vq} \frac{i\vk}{k^2} \delta_L(\vk) ,
\end{equation}

This formalism makes it particularly easy to include redshift space
distortions.
In this work we adopt the standard ``plane-parallel'' or ``distant-observer''
approximation, in which the line-of-sight direction to each object is taken to
be the fixed direction $\hat{z}$. This has been shown to be a good
approximation at the level of current observational error bars
(e.g., Figure 10 of \citealt{SamPerRac12} or Figure 8 of \citealt{YooSel14}).
Under this assumption, the position of an object located at true comoving
position $\vx$, will be mis-identified due to its peculiar velocity along
the line-of-sight, as
\begin{equation}
    \mathbf{s} = \vx + \frac{\hat{z} \cdot\mathbf{v}(\vx)}{aH} \hat{z} .
\end{equation}
Thus including redshift-space distortions requires only a simple additive
offset of the displacement field.
The peculiar velocity of a fluid element, labeled by its Lagrangian coordinate
$\vq$, is $\mathbf{v}(\vq) = a \dot{\vPsi}(\vq)$ so in redshift space the
apparent displacement of the fluid element is
\begin{equation}
    \vPsi^s = \vPsi + \frac{\hat{z} \cdot \dot{\vPsi}}{H} \hat{z} .
\end{equation}
To a good approximation the time dependence of the $n$th order term
in Eq.~(\ref{eq:psin}) is given by $\vPsi^{(n)} \propto D^n$.  Therefore
$\dot{\vPsi}^{(n)} = nHf \vPsi^{(n)}$, where $f = d\log D/d\log a$ is the
growth rate, often approximated as $f \approx \Omega_m^{0.55}$.
Thus, within the Zel'dovich approximation, the mapping to redshift space
is achieved via the matrix
\begin{equation}
 \Psi_i \to \Psi_i^s = (\delta_{ij} + f \hz_i \hz_j) \Psi_j
\end{equation}
which simply multiplies the $z$-component of the vector by $1+f$.

Finally we must consider biased tracers of the density field.
We begin by considering a local Lagrangian bias model, which posits that
the locations of discrete tracers at some late time are determined by the
overdensities in the initial matter density field, specifically:
\begin{equation}
    \rho_X(\vq) = \bar{\rho}_X F[\delta_R(\vq)] .
\end{equation}
Here $\bar{\rho}_X$ is the mean comoving number density of our tracer $X$ and
the function $F(\delta)$ is called the Lagrangian bias function.
\citet{Mat11} provides an extensive discussion of non-local Lagrangian bias.

\begin{figure}
\begin{center}
\resizebox{3.3in}{!}{\includegraphics{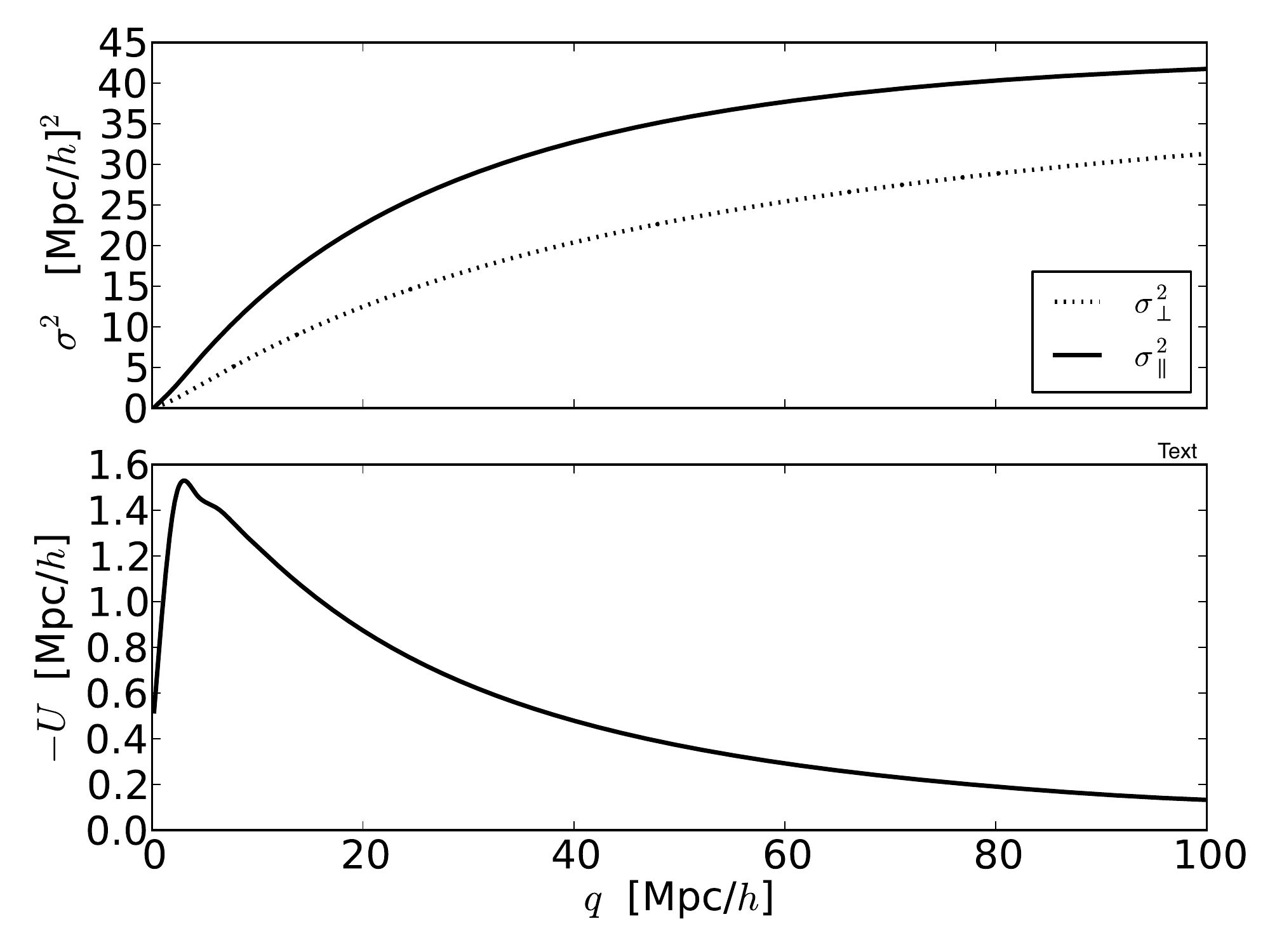}}
\end{center}
\caption{The 2-point functions (Eqs.~\ref{eqn:qf0}-\ref{eqn:qf}) which
enter into the computations.  The upper panel shows the dispersions,
$\sigma^2_\perp=2(\sigma_\eta^2-\eta_\perp)$ and
$\sigma^2_\parallel=2(\sigma_\eta^2-\eta_\parallel)$,
as a function of $q$ while the lower panel shows the mean velocity, $-U(q)$.}
\label{fig:qf}
\end{figure}

The correlation function within the Zel'dovich approximation then follows
by elementary manipulations
\citep{BonCou88,FisNus96,Mat08a,Mat08b,CLPT,WanReiWhi13,Tas14b}.  We
begin by writing
\begin{equation}
    1 + \delta_X(\vx)
    =\int d^3q~ F[\delta_R(\vq)] \delta_D\left[\vx - \vq - \vPsi(\vq)\right]
    \ .
\end{equation}
We now replace the delta function with its Fourier representation, and also
introduce the Fourier\footnote{An alternative route to the same final
expressions is to expand $F(\delta_R)$ in powers of $\delta_R$ and use the
properties of Gaussian integrals.  We will use the Fourier methodology since
it was also used in \citet{Mat08b,CLPT,WanReiWhi13}.} transform $F(\lam)$
of $F(\delta)$, so the expression for $1+\delta_X$ becomes
\begin{align}
  1 + \delta_X(\vx)
  &= \int d^3q~ F[\delta_R(\vq)] \int \frac{d^3k}{(2\pi)^3}
  \ e^{i\vk\cdot[\vx-\vq-\vPsi(\vq)]} \\
  &= \int d^3q \int \frac{d^3k}{(2\pi)^3} \int \frac{d\lam}{2\pi}\ F(\lam)
  \ e^{i\left\{\lam \delta_R(\vq) + \vk\cdot[\vx-\vq-\vPsi(\vq)]\right\}} .
\end{align}
The 2-point correlation function
$\xi_X(\vr) = \langle \delta_X(\vx_1) \delta_X(\vx_2) \rangle$
for the biased tracer $X$ is then given by
\begin{eqnarray}
    1 + \xi_X(\vr)
        &=& \int d^3q_1\ d^3q_2
            \int \frac{d^3k_1}{(2\pi)^3} \frac{d^3k_2}{(2\pi)^3}
        \ e^{i\vk_1\cdot(\vx_1-\vq_1)} e^{i\vk_2\cdot(\vx_2-\vq_2)}
        \nonumber \\
   &\times& \int \frac{d\lam_1}{2\pi} \frac{d\lam_2}{2\pi}
   \ F(\lam_1) F(\lam_2)
   \left\langle e^{i[\lam_1 \delta_1 + \lam_2 \delta_2
   - \vk_1\cdot\vPsi_1 - \vk_2\cdot\vPsi_2]} \right\rangle
   \quad ,
\end{eqnarray}
where $\delta_a \equiv \delta_R(\vq_a)$, $\vPsi_a \equiv \vPsi(\vq_a)$, and
$\vr = \vx_2 - \vx_1$.
By statistical homogeneity, the expectation value above depends only on the
difference in Lagrangian coordinates, $\vq = \vq_2 - \vq_1$.
The change of variables
$\{\vq_1,\vq_2\} \to \left\{\vq,\vec{Q} = (\vq_1+\vq_2)/2\right\}$
then leads to
\begin{equation}
    \label{eq:1+xi}
    1 + \xi_X(\vr)
        = \int d^3q
            \int \frac{d^3k}{(2\pi)^3} e^{i\vk\cdot(\vq-\vr)}
            \int \frac{d\lam_1}{2\pi} \frac{d\lam_2}{2\pi}
            \ F_1 F_2\,K(\vq,\vk,\lam_1,\lam_2) ,
\end{equation}
where we have defined
\begin{equation}
    K(\vq,\vk,\lam_1,\lam_2) = \left\langle
    e^{i(\lam_1 \delta_1 + \lam_2 \delta_2 + \vk\cdot\vDelta)} \right\rangle ,
\label{eqn:K}
\end{equation}
and $\Delta \equiv \vPsi_2 - \vPsi_1$.  This expression, derived in
\citet{CLPT}, is the exact configuration space analog of Eq.~(9) in
\citet{Mat08b} and is analogous to the power spectrum derived in
\citet{FisNus96}.

We can expand the expectation value in Eq.~(\ref{eqn:K}) in terms of cumulants.
Since $\delta_R(\vq)$ and $\vPsi$ are Gaussian only the second cumulant is
non-zero
\begin{eqnarray}
  \left\langle \left(\lam_1 \delta_1 + \lam_2 \delta_2
  + \vk\cdot\vDelta\right)^2 \right\rangle_c
  &=& (\lam_1^2 + \lam_2^2) \sigma_R^2 + A_{ij} k_i k_j \nonumber \\
  &+& 2 \lam_1 \lam_2 \xi_R + 2(\lam_1 + \lam_2) U_i k_i ,
\end{eqnarray}
where we have defined
\begin{align}
    \sigma_R^2 &= \langle \delta_1^2 \rangle_c
                = \langle \delta_2^2 \rangle_c , &
    \xi_R(\vq) &= \langle \delta_1 \delta_2 \rangle_c , \\
    A_{ij}(\vq)&= \langle \Delta_i \Delta_j \rangle_c , &
    U_i(\vq)   &= \langle \delta_1 \Delta_i \rangle_c
                = \langle \delta_2 \Delta_i \rangle_c .
\end{align}
Eq. (\ref{eqn:K}) then evaluates to
\begin{equation}
    K = \exp\left[-\frac{1}{2} (\lam_1^2 + \lam_2^2) \sigma_R^2 -
    \frac{1}{2} A_{ij} k_i k_j - \lam_1 \lam_2 \xi_R -
    (\lam_1 + \lam_2) U_i k_i \right] .
\label{eqn:K_ZA}
\end{equation}
This expression is exact, within the Zel'dovich approximation.
The quantity $\sigma_R^2$ is simply the variance of the smoothed linear density
field, while $\xi_R(\vq) = \langle \delta_R(\vq_1) \delta_R(\vq_2) \rangle$ is
the corresponding smoothed linear correlation function.  The matrix $A_{ij}$
may be decomposed as
\begin{eqnarray}
  A_{ij}(\vq) &=& 2\left[\sigma_\eta^2 - \eta_\perp(q)\right] \delta_{ij}
               + 2\left[\eta_\perp(q) - \eta_\parallel(q)\right] \hq_i \hq_j ,\\
  &=& \sigma_\perp^2\delta_{ij}
   + \left[\sigma_\parallel^2-\sigma_\perp^2\right]\hq_i\hq_j
\label{eqn:Aijdef}
\end{eqnarray}
where $\sigma_\eta^2 \equiv \frac{1}{3} \langle |\vPsi|^2 \rangle$ is the 1-D
dispersion of the displacement field, and $\eta_\parallel$ and $\eta_\perp$ are
the transverse and longitudinal components of the Lagrangian 2-point function,
$\eta_{ij}(\vq) = \left\langle \Psi_i(\vq_1) \Psi_j(\vq_2) \right\rangle$.
The vector $U_i(\vq)=U(q)\,\hq_i$ is the cross-correlation between the linear
density field and the Lagrangian displacement field.
In the Zel'dovich approximation these quantities are given by
\begin{gather}
  \sigma_\eta^2 = \frac{1}{6\pi^2} \int_0^\infty dk~ P_L(k) , \label{eqn:qf0}
  \\
  \eta_\perp(q) = \frac{1}{2\pi^2} \int_0^\infty dk~ P_L(k)~
  \frac{j_1(kq)}{kq} , \\
  \eta_\parallel(q) = \frac{1}{2\pi^2} \int_0^\infty dk~ P_L(k)~
  \left[j_0(kq) - 2 \frac{j_1(kq)}{kq}\right] , \\
  U(q) = -\frac{1}{2\pi^2} \int_0^\infty dk~ k P_L(k)~ j_1(kq) .
\label{eqn:qf}
\end{gather}
and shown in Fig.~\ref{fig:qf}.
Up to factors of 2 and $f$, these expressions are identical to the Eulerian
velocity correlators in linear theory \citep[e.g.][]{Gor88,Fis95,ReiWhi11},
which is not surprising since $\mathbf{v}_L = f \Psi$ in the Zel'dovich
approximation.
It is also useful to define
$\sigma_{12}^2=2[\sigma_\eta^2-\mu^2\eta_\parallel-(1-\mu^2)\eta_\perp]$,
the pairwise velocity dispersion at an angle $\mu$ to the line-of-sight
with the line-of-sight and perpendicular components $\sigma_\parallel^2$
and $\sigma_\perp^2$.

\section{Correlation function -- matter}
\label{sec:matter}

For the unbiased case (i.e.~the matter field) we can write our expression
for $\xi^{(ZA)}$ in closed form by carrying out the Gaussian integral
\begin{align}
    1 + \xi^{(ZA)}(\vr)
        &= \int d^3q \int \frac{d^3k}{(2\pi)^3}
  e^{i\vk\cdot(\vq-\vr)} e^{-\frac{1}{2} A_{ij} k_i k_j} \\
        &= \int \frac{d^3q}{(2\pi)^{3/2} |A|^{1/2}}
  e^{-\frac{1}{2} (\vr-\vq)^\top \mathbf{A}^{-1} (\vr-\vq)} ,
\end{align}
Further discussion of this expression, and the approach to linear theory,
can be found in \citet{CLPT}.

Evaluation of $\xi^{(ZA)}$ involves a numerical integral of a Gaussian
function.  The inversion of $A_{ij}$ which is required can be done
analytically by use of the Sherman-Morrison formula which states that for
matrices $M$ and vectors $b$ and $c$,
\begin{equation}
  \left( M + bc^T\right)^{-1} = M^{-1} - \frac{M^{-1}bc^TM^{-1}}{1+c^TM^{-1}b}
  \quad .
\end{equation}
Writing $A_{ij}=F\delta_{ij}+G\ \widehat{q}_i\widehat{q}_j$
(see Eq.~\ref{eqn:Aijdef}) we have
\begin{eqnarray}
  A^{-1}_{ij} &=& \frac{\delta_{ij}}{F} -
  \frac{G\ \widehat{q}_i\widehat{q}_j}{F(F+G)} \\
  &=& \frac{\delta_{ij}}{\sigma_\perp^2} +
      \frac{\sigma_\perp^2-\sigma_\parallel^2}
           {\sigma_\parallel^2\sigma_\perp^2}\hq_i\hq_j
\end{eqnarray}
where $F=\sigma_\perp^2$, $G=\sigma_\parallel^2-\sigma_\perp^2$ and the
combination $F+G=\sigma_\parallel^2$.
To compute the determinant we make use of det$(cM)=c^3{\rm det} M$ for
scalar $c$ and $3\times 3$ matrix $M$ and that det$(I+uv^T)=1+u^Tv$.  Then
\begin{equation}
  {\rm det}\,A = \left(\sigma_\perp^2\sigma_\parallel\right)^2
\end{equation}
as expected.
The integrand can thus be expressed analytically in terms of the 2-point
functions defined previously and evaluated by simple quadratures\footnote{We
use the midpoint rule in $|\vq-\vr|$ and Gauss-Legendre integration in
$\widehat{q}\cdot\widehat{r}$.} in $\mathbf{x}=\mathbf{q}-\mathbf{r}$.
The integral is dominated by $\mathbf{q}\approx\mathbf{r}$.
By tabulating the functions $\eta_i(q)$ and $U(q)$ in advance these integrals
can be performed very quickly.
In redshift space we replace
\begin{align}
 U_i & \to U_i^s = (\delta_{ij} + f \hz_i \hz_j) U_j , \\
 A_{ij} & \to A_{ij}^s =
 (\delta_{ik} + f \hz_i \hz_k) (\delta_{jl} + f \hz_j \hz_l) A_{kl} .
\end{align}
This corresponds simply to dividing the $z$-components of the inverse of
$A$ by $1+f$ and multiplying $U_z$ by $1+f$.

\section{Perturbative expansion for biased tracers}
\label{sec:biased}

Returning to the case of biased tracers, consider again Eq. (\ref{eqn:K_ZA}).
In the unbiased case the $\vk$ integration in Eq. (\ref{eq:1+xi}) took the
form of a Gaussian integral, which we carried out analytically.
In the biased case, we can achieve the same thing if we first partially
expand Eq. (\ref{eqn:K_ZA}) as
\begin{eqnarray}
    K &=& e^{-\frac{1}{2}(\lam_1^2 + \lam_2^2) \sigma_R^2}
    e^{-\frac{1}{2} \vk^T A \vk} \left[1 - (\lam_1 + \lam_2) U_i k_i
    - \lam_1 \lam_2 \xi_R \right.  \nonumber \\
   &+& \frac{1}{2} (\lam_1+\lam_2)^2 U_i U_j k_i k_j
    + \lam_1 \lam_2 (\lam_1+\lam_2) \xi_R U_i k_i
    + \frac{1}{2} \lam_1^2 \lam_2^2 \xi_R^2 \nonumber \\
   &-& \frac{\lam_1^3\lam_2^3}{3!}\xi_R^3
            - \frac{1}{2}\lam_1\lam_2(\lam_1+\lam_2)^2U_iU_jk_ik_j\xi_R
       \nonumber \\
   &-& \frac{1}{2}\lam_1^2\lam_2^2(\lam_1+\lam_2)U_ik_i\xi_R^2
            - \frac{(\lam_1+\lam_2)^3}{3!}U_iU_jU_nk_ik_jk_n \nonumber \\
   &+& \left. \frac{\lam_1^4\lam_2^4}{4!}\xi_R^4 +  \cdots \right]
\label{eqn:K_expanded}
\end{eqnarray}
We may justify this choice of expansion by noting that both $\xi_R(\vq)$ and
$U_i(\vq)$ vanish in the large-scale limit $|\vq| \to \infty$, while
$\sigma_R^2$ and $A_{ij}(\vq)$ approach non-zero values.
At this stage we note one difference between doing this expansion within
Lagrangian perturbation theory \citep[e.g.][]{Mat08b,OkaTarMat11,CLPT} and
performing it within the context of the Zel'dovich expansion.
In the Zel'dovich approximation all the terms are just multiples of 2-point
functions and we can go to arbitrarily high order without the need to evaluate
any high dimensional mode coupling integrals or numerically difficult terms.
Here we have gone to cubic order in the 2-point function and indicated how the
quartic terms appear.  We shall show later that the expansion seems to be
converging quickly.

\begin{figure*}
\begin{center}
\resizebox{6.5in}{!}{\includegraphics{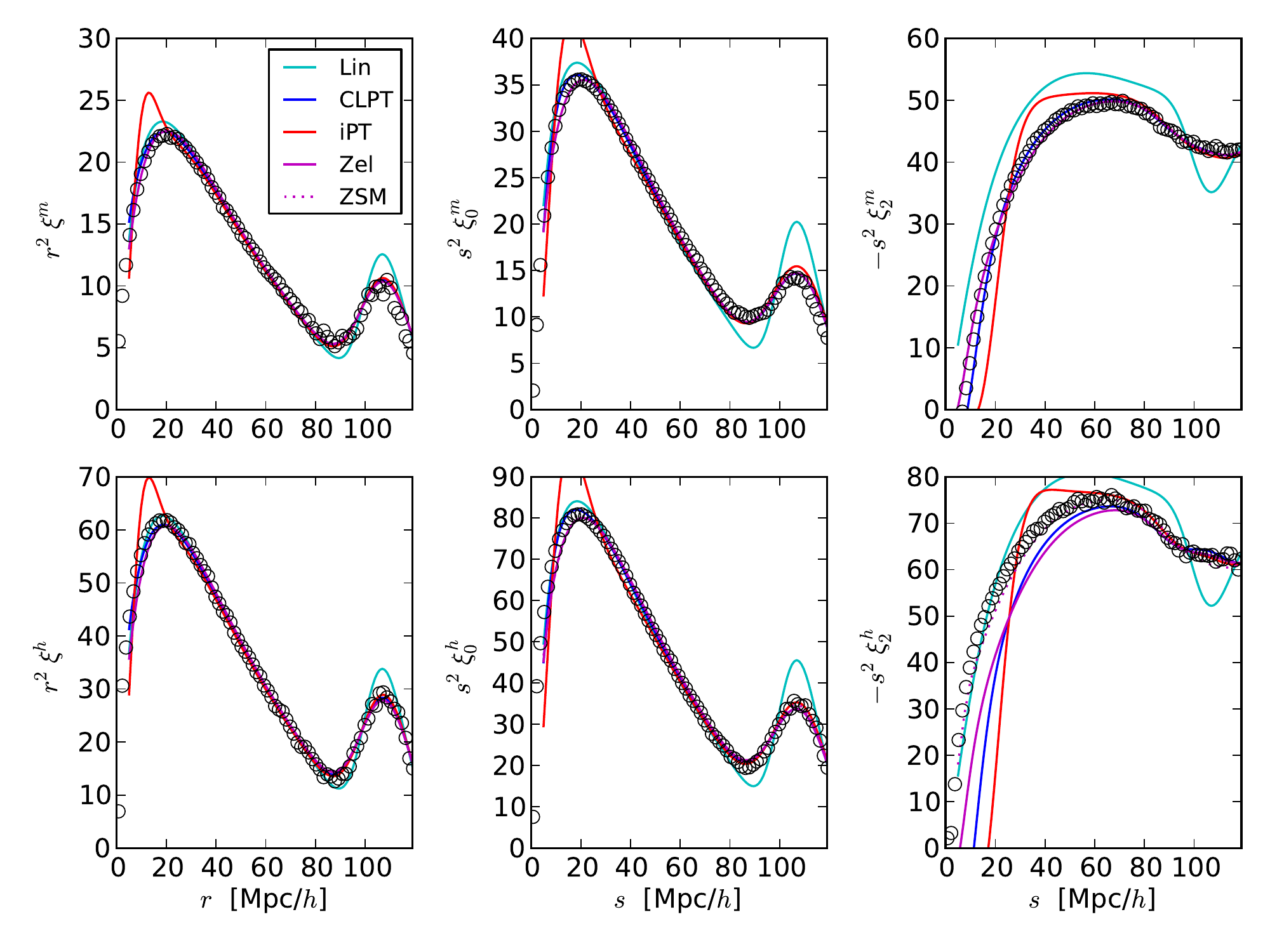}}
\end{center}
\caption{A comparison of different Lagrangian perturbative schemes to
N-body results for the 2-point function, $\xi$, in real- and redshift-space.
The results are for the cosmology used in \citet{CLPT}, which has
$\Omega_m=0.274$, at $z=0.55$.  The points show the average over 20
realizations of the N-body simulation while the lines show the analytic
approximations: linear theory (cyan);
convolution Lagrangian perturbation theory \citep[CLPT;][blue]{CLPT};
integrated perturbation theory \citep[iPT;][red]{Mat08b};
the Zel'dovich approximation (magenta) and the Zel'dovich streaming model
(ZSM; dotted magenta).
The upper row is for the matter, while the lower row is for biased tracers
with $b\simeq 1.6$ (friends-of-friends halos with
$12.785<\log_{10}M_h/(h^{-1}M_\odot)<13.085$).  From left to right the
columns are the real-space correlation function, the redshift-space monopole
and the redshift-space quadrupole, all multiplied by $r^2$ to allow a linear
$y$-axis.  Note that the Zel'dovich approximation provides a good fit to the
N-body data on large scales for all but the quadrupole moment of the
redshift-space halo correlation function (lower right panel).  The fact that
the iPT, CLPT and Zel'dovich lines are almost indistinguishable on large
scales shows the good convergence of Lagrangian perturbation theory schemes.}
\label{fig:cmp_zel}
\end{figure*}

The $\lam_1$ and $\lam_2$ integrations give
$(-i)^n\left\langle F^{(n)}\right\rangle$,
the expectation value of the $n$th derivative of the Lagrangian bias function
$F(\delta)$ \citep{Mat08b}.  In order to make the expressions more readable
we shall write $b_n=\left\langle F^{(n)}\right\rangle$.  Then
\begin{eqnarray}
  L(\vq,\vk) &\equiv&
  \int \frac{d\lam_1}{2\pi} \frac{d\lam_2}{2\pi}
  \ F(\lam_1)F(\lam_2)\,K(\vq,\vk,\lam_1,\lam_2) \\
  &=& e^{-\frac{1}{2} A_{ij} k_i k_j} \left[
  1 + b_1^2 \xi_R + 2i b_1 U_i k_i + \frac{1}{2} b_2^2 \xi_R^2 \right.
  \nonumber \\
  &-& (b_2 + b_1^2) U_i U_j k_i k_j + 2i b_1 b_2 \xi_R U_i k_i  \nonumber \\
  &+& \frac{b_3^2}{3!}\xi_R^3 - (b_1b_3+b_2^2)U_iU_jk_ik_j\xi_R \nonumber \\
  &+& ib_2b_3U_ik_i\xi_R^2 - i(b_1b_2+\frac{1}{3}b_3)U_iU_jU_nk_ik_jk_n
      \nonumber \\
  &+& \left. \cdots \right] .
  \label{eqn:L_ZA}
\end{eqnarray}
The $\vk$ integration reduces to a series of multi-variate Gaussian integrals
which can be done with the formulae in \citet{CLPT}.
In the end we obtain
\begin{eqnarray}
  M(\vr,\vq) &\equiv&
  \int \frac{d^3k}{(2\pi)^3}~ e^{i\vk\cdot(\vq-\vr)} L(\vq,\vk) \\
  &=& \frac{1}{(2\pi)^{3/2} |A|^{1/2}}
  \ e^{-\frac{1}{2} (\vr-\vq)^T \mathbf{A}^{-1} (\vr-\vq)}
  \left[ 1 + b_1^2 \xi_R \right. \nonumber \\
  &-& 2 b_1 U_i g_i + \frac{1}{2} b_2^2 \xi_R^2
   - (b_2 + b_1^2) U_i U_j G_{ij} \nonumber \\
  &-& 2b_1 b_2 \xi_R U_i g_i + \frac{b_3^2}{3!}\xi_R^3
   - (b_1b_3+b_2^2)G_{ij}U_iU_j\xi_R \nonumber \\
  &-& b_2b_3U_ig_i\xi_R^2 + (b_1b_2+\frac{1}{3}b_3)\Gamma_{ijn}U_iU_jU_n
     \nonumber \\
  &+& \left. \cdots \right] ,
\label{eqn:M_ZA}
\end{eqnarray}
where
\begin{eqnarray}
  g_i &\equiv& (A^{-1})_{ij}(q-r)_j \nonumber \\
  G_{ij}&\equiv& (A^{-1})_{ij} - g_i g_j \nonumber \\
  \Gamma_{ijk}&\equiv& (A^{-1})_{ij} g_k + (A^{-1})_{ki} g_j
                     + (A^{-1})_{jk} g_i - g_i g_j g_k
\label{eq:g_i}
\end{eqnarray}
so that
\begin{equation}
  \Gamma_{ijn}U_iU_jU_n = 3\left(U_i G_{ij} U_j\right)\left(U_ng_n\right)
  + 2\left(U_ig_i\right)^3
\end{equation}
Our final expression for the correlation function is
\begin{equation}
    1 + \xi_X(\vr) = \int d^3q~ M(\vr,\vq) .
\label{eq:Z_ZA}
\end{equation}
The remaining integration over $\vq$ must be performed numerically as before.

One can treat the $b_n$ as fitting parameters, or attempt to compute them from
a bias model.  One such model is the peak-background split, which begins with
the unconditional multiplicity function
\begin{equation}
 \nu f(\nu)\,d\nu = \frac{M}{2\bar{\rho}}\ \frac{dn}{dM}\, dM
\end{equation}
which can be fit with
\begin{equation}
\label{eq:stmass}
  \nu f(\nu) \propto \left(1 + \frac{1}{(a\nu^{2})^{p}}\right)
  \left(\frac{a\nu^2}{2}\right)^{1/2}
  \exp\left(- \frac{a\nu^{2}}{2}\right)
\end{equation}
where $a=1$, $p=0$ gives the Press-Schecter mass function \citep{PS},
while $a=0.707$, $p=0.3$ yields the Sheth-Tormen mass function \citep{SheTor99}.
Within the assumption of the peak-background split, the conditional
multiplicity function is given by the substitution,
\begin{equation}
  \nu \rightarrow \nu \left(1 - \frac{\delta}{\delta_{c}}\right) \,\,,
\end{equation}
where $\delta$ is the background density and $\delta_c\simeq 1.686$ is the
critical overdensity for collapse.
The Lagrangian bias parameters then follow from Taylor expanding the
(appropriately normalized) conditional multiplicity function as a function
of $\delta$, yielding $b_n=[\nu f(\nu)]^{-1} d^n/d\delta^n[\nu f(\nu)]$ or
\begin{equation}
  b_1(\nu) = \frac{1}{\delta_c} \left[ a\nu^{2} - 1
  + \frac{2p}{1 + (a\nu^{2})^{p}}\right] \,,
\label{eqn:blag1}
\end{equation}
\begin{equation}
  b_2(\nu) = \frac{1}{\delta_c^2} \left[a^2 \nu^4 -3 a \nu^2 +
       \frac{2p(2a\nu^2 + 2p - 1)}{1 + (a\nu^{2})^{p}} \right]\,,
\label{eqn:blag2}
\end{equation}
and
\begin{eqnarray}
  b_3(\nu) &=& \frac{1}{\delta_c^3} \left[a^3 \nu^6 -6 a^2 \nu^4 +
      3a\nu^2 + \right. \nonumber \\
  &+& \left. \frac{2p(3a^2\nu^4+6a\nu^2(p-1)+4p^2-1)}{1+(a\nu^{2})^{p}}
       \right]\,\,.
\label{eqn:blag3}
\end{eqnarray}
For the halo sample shown in Fig.~\ref{fig:cmp_zel} we have $\nu\simeq 1.6$,
$b_1=0.64$, $b_2=-0.45$ and $b_3=-1.63$ from the Sheth-Tormen mass function
and we have used these values in the relevant figures.
There is some evidence \citep{Bal12} that simplistic bias models such as
the above are not quantitatively accurate when compared to N-body simulations.
On large scales the level of agreement is quite insensitive to the value
assumed for $b_{n\ge 2}$ as long as $|b_{n\ge 2}|$ is not much larger than
$b_1$ because the terms involving $b_{n\ge 2}$ are numerically small compared
to the $b_1$ terms.
Thus assuming the peak-background split value for $b_{n\ge 2}$ is perfectly
adequate and in matching the Zel'dovich theory to observations there is only
one free parameter ($\nu$ or $b_1$).  This is very well constrained by the
overall amplitude of $\xi$.

\begin{figure*}
\begin{center}
\resizebox{6.5in}{!}{\includegraphics{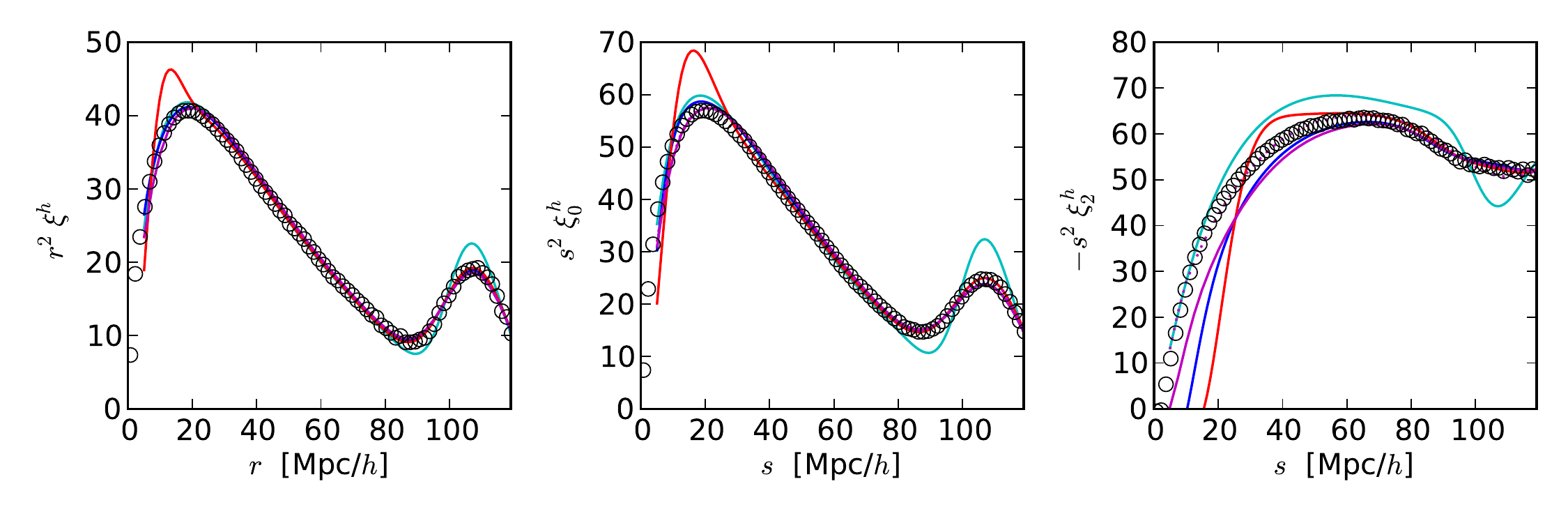}}
\end{center}
\caption{As for Fig.~\ref{fig:cmp_zel} except for a different halo mass
range.  The panels and line types are the same as the lower three panels
of Fig.~\ref{fig:cmp_zel} but for friends-of-friends halos with
$12.182<\log_{10}M_h/(h^{-1}M_\odot)<12.483$ with large-scale bias
$b\simeq 1.3$.}
\label{fig:three_panel}
\end{figure*}

\section{Results}
\label{sec:results}

Figs.~\ref{fig:cmp_zel} and \ref{fig:three_panel} compare the predictions of
the Zel'dovich approximation to a number of other Lagrangian perturbation
theory schemes and to the clustering of halos in an N-body simulation.
The agreement between the predictions of the Zel'dovich approximation and
N-body simulations is very good on large scales.
For the real-space correlation function and the redshift-space monopole the
agreement extends down to $10-20\,h^{-1}$Mpc for both the matter and halos.
However, the theory does much less well for the halo quadrupole (a problem
shared with all of the local Lagrangian bias schemes shown in
Fig.~\ref{fig:cmp_zel}).  One major difference between the halo calculation
and the matter calculation is the Taylor series expansion of the bias terms.

While the perturbative schemes plotted in Fig.~\ref{fig:cmp_zel} stop at
$\mathcal{O}(P_L^2)$, we have extended the Zel'dovich calculation one
additional order to see if the higher order terms could contribute to the
quadrupole on intermediate and small scales.
We find that the terms cubic in $\xi_R$ and $U$ contribute negligibly
to the real-space correlation function and the monopole and quadrupole moments
of the redshift-space correlation function for $r>10\,h^{-1}$Mpc.
Since the $\mathcal{O}(P_L^3)$ terms do not contribute significantly
to any of the statistics for this halo sample it appears that the Taylor
series expansion is not the source of the discrepancy.
This also suggests that truncating the expansion at $\mathcal{O}(P_L^2)$
is a good approximation and that the expansion is converging.
Henceforth we shall drop the $\mathcal{O}(P_L^3)$ terms.

\begin{figure}
\begin{center}
\resizebox{3.3in}{!}{\includegraphics{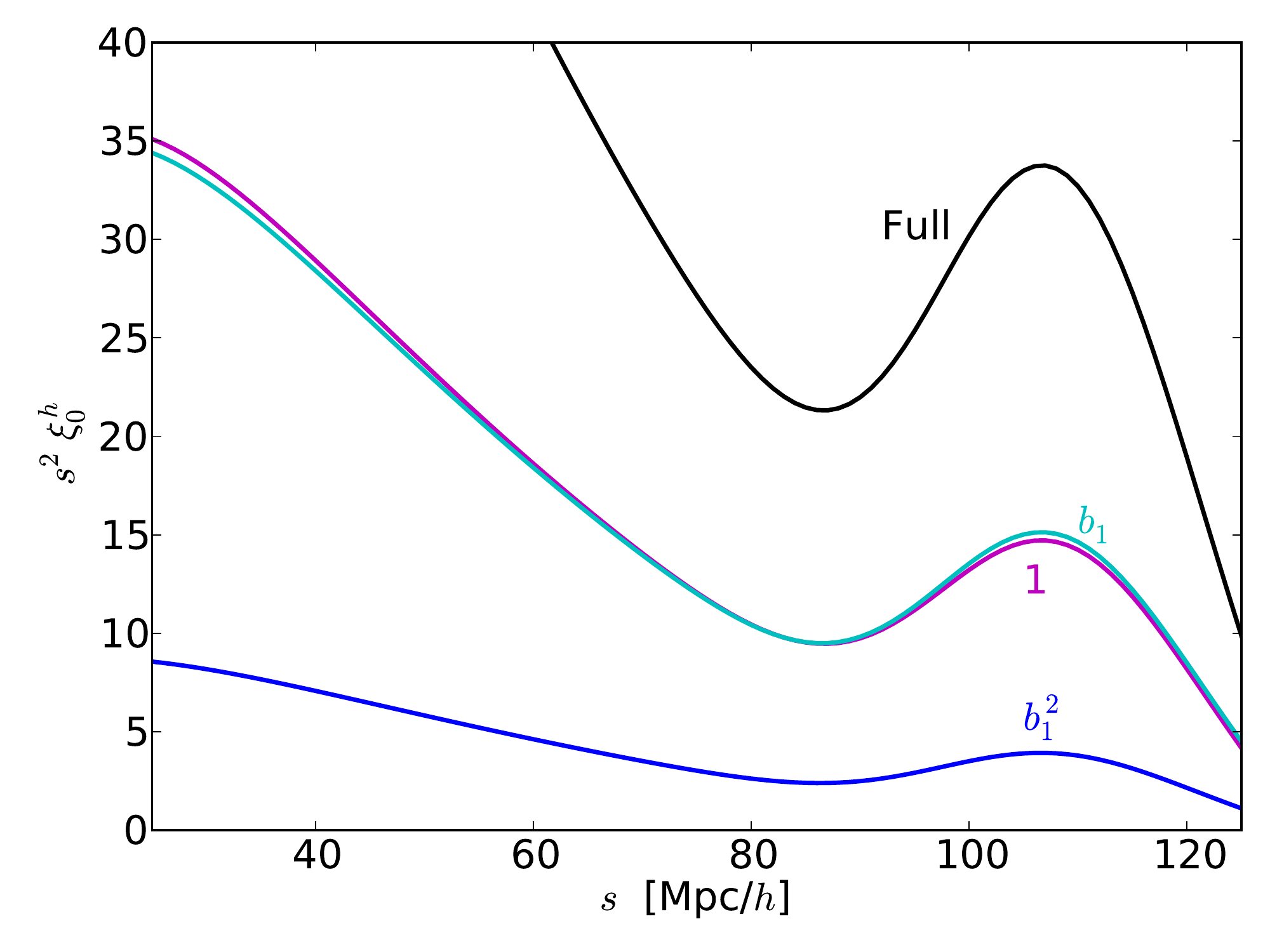}}
\resizebox{3.3in}{!}{\includegraphics{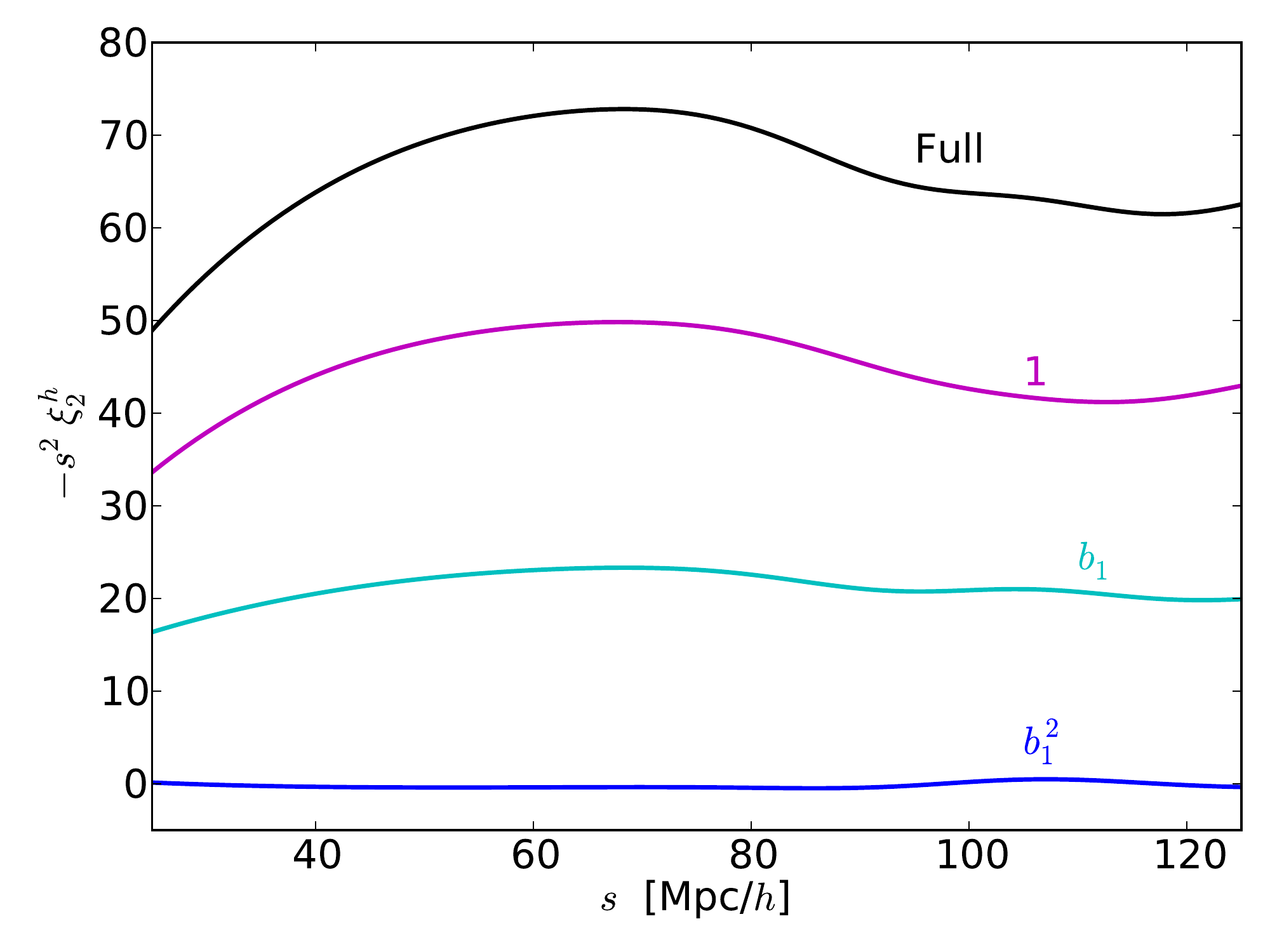}}
\end{center}
\caption{(Top) The peak in $s^2\xi_0^h(s)$ near $110\,h^{-1}$Mpc and the
contributions from the various terms.  The black line shows the full
Zel'dovich prediction, which is is indistinguishable from the prediction
with $b_2=0$.  The magenta line shows the contribution from the ``$1$'' term
in Eq.~(\ref{eqn:M_ZA}).  The cyan line shows the contribution from the term
linear in $b_1$ and the blue line the term quadratic in $b_1$.  (Bottom) The
same for the quadrupole, $-s^2\xi_2^h(s)$.}
\label{fig:monopole_peak}
\end{figure}

The contribution to the correlation functions is not shared equally among
the terms.  Working above $r=10\,h^{-1}$Mpc we find that the real-space
correlation function is dominated by the $U_ig_i$, ``$1$'' and $\xi_R$ terms
in the square brackets of Eq.~(\ref{eqn:M_ZA}).
The redshift-space monopole is dominated by the same three terms.
Fig.~\ref{fig:monopole_peak} shows how the different terms contribute to
the final peak in $s^2\xi_0^h(s)$ near $110\,h^{-1}$Mpc and to the quadrupole.
Above $50\,h^{-1}$Mpc the terms ``$1$'' and $U_ig_i$ in the square brackets
in Eq.~(\ref{eqn:M_ZA}) contribute the vast majority of the total
quadrupole signal, with approximately equal contributions.
By $20\,h^{-1}$Mpc the other terms contribute about 10 per cent of the total,
with the $U_iU_jG_{ij}$, $U_ig_i\xi_R$ and $\xi_R$ terms contributing the
remainder in decreasing order of importance (the $\xi_R^2$ term provides a
negligible contribution).

\begin{figure}
\begin{center}
\resizebox{3.3in}{!}{\includegraphics{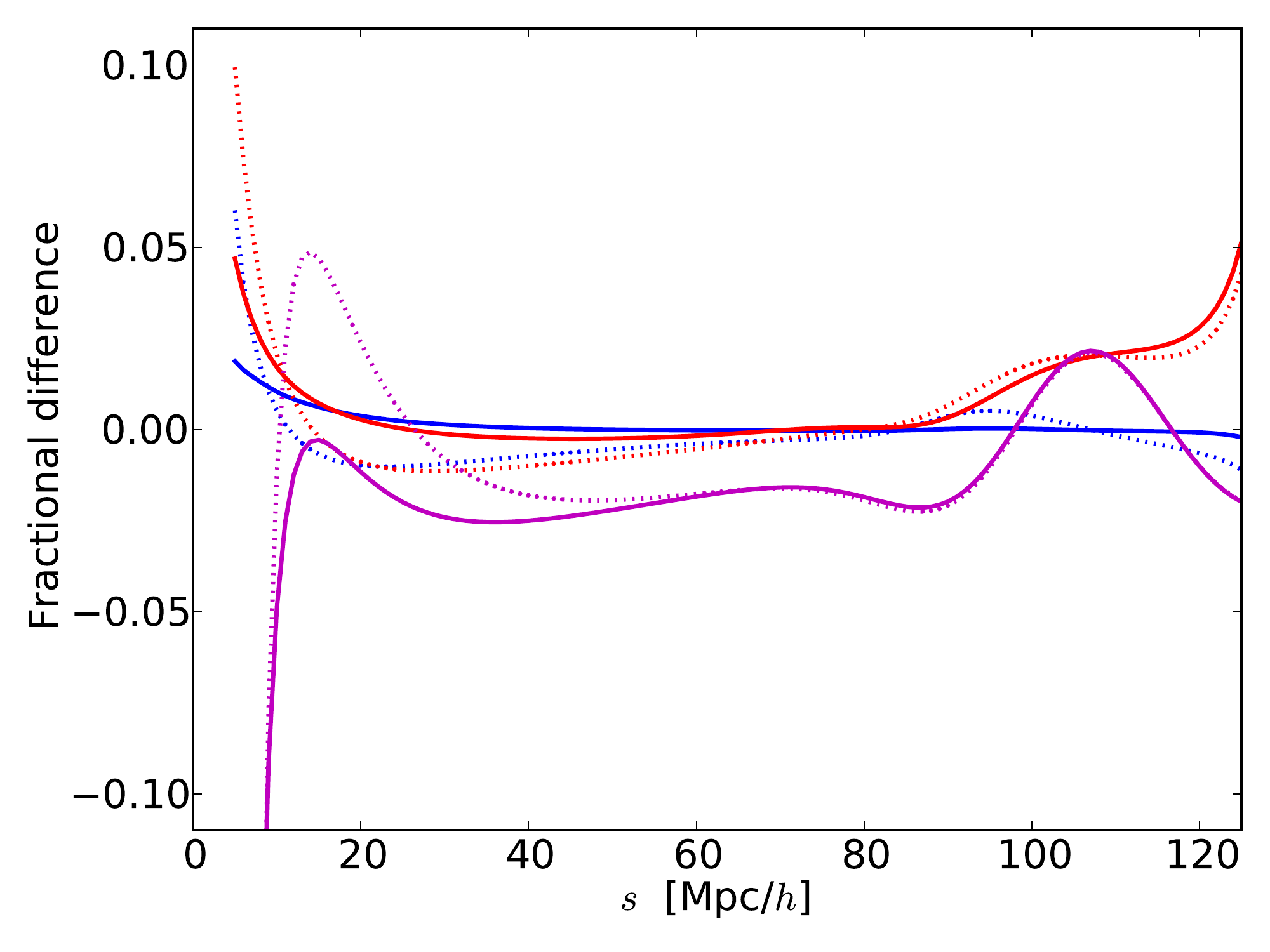}}
\end{center}
\caption{The fractional difference between the full Zel'dovich calculation
for the halo real-space correlation function (blue), redshift-space monopole
(red) and quadrupole (magenta) and the `constant bias' approximations:
$(1+b_1)^2\xi^m$,
$(1+b_1)^2(1+[2/3]\beta+[1/5]\beta^2)/(1+[2/3]f+[1]5/f^2)\xi_0^m$ and
$[b/3+f/7]/[1/3+f/7]\xi_2^m$.
Here $\beta\equiv f/(1+b_1)$ and $f\approx 0.744$ for this cosmology and
redshift.
The solid lines include all of the terms, while the dotted lines show the
computation with $b_2=0$.}
\label{fig:plot_pieces}
\end{figure}

Fig.~\ref{fig:plot_pieces} shows the degree to which the $b_1$ and $b_2$ terms
depend on scale differently than the matter terms.
In all cases, above $10\,h^{-1}$Mpc the level of scale-dependence is quite
small.  The actual shape of the real-space correlation function and the
redshift-space monopole correlation function differ at the ten per cent level
near the peak, but this difference is due to the impact of redshift space
distortions on the correlation function and not due to scale-dependent bias
in the sense that we mean it here.
Note that this relative scale-independence does not need to hold in Fourier
space.  As a trivial example, $P(k)=b^2P_L(k)+N(k)$ has a scale-independent
configuration-space bias if the transform of $N(k)$ is arbitrarily small on
the scales of interest.  Taking $N(k)=\bar{n}^{-1}$,
$N(k)\propto\exp[-k^2R^2]$ or the convolution of two halo profiles can satisfy
this criterion.
Depending on the size of $N(k)/P_L(k)$ this could lead to a large
(but smoothly varying) scale-dependent bias in Fourier space
\citep[see e.g.~the discussion in][]{SchWhi06}.
Conversely, one finds that the Fourier transform of the Zel'dovich
correlation function has almost no power beyond $k\sim\sigma_\eta^{-1}$.
To use the Zel'dovich approximation in Fourier space requires the addition
of other terms which provide the missing power but affect the correlation
function only at small scales.

Interestingly, the Zel'dovich approximation predicts that halos which are
locally biased in Lagrangian space will have approximately the same small-scale
quadrupole to large-scale quadrupole ratio as the matter, while the halos in
N-body simulations display a significantly larger small-scale quadrupole when
scaled to the same large-scale quadrupole.
The term involving $b_2$ gives the desired increase in small-scale quadrupole,
though at too small an amplitude.
Making $b_2$ large and positive helps slightly, but the quadrupole still has
the wrong overall shape.

Some authors \citep[e.g.][]{MelPelSha94} have suggested that the
Zel'dovich approximation performs better when small-scale power is filtered
out of the linear theory power spectrum.  We have tested this by Gaussian
filtering the input $P_L(k)$ on scales $1-5\,h^{-1}$Mpc.  We find that none
of these scales improves the agreement of the quadrupole of the redshift-space
correlation function on smaller scales.

At this point it is unclear whether the discrepancy above is due to our
assumption of local Lagrangian bias or the Zel'dovich dynamics predicting the
wrong velocity field for halos.
To explore this issue further, we have run another set of 8 simulations with
a simplified set-up.
Each simulation started with initial conditions generated with the Zel'dovich
approximation at $z_{ic}=67$ (where the rms displacement was about 10 per cent
of the inter-particle spacing).  To ensure numerical convergence we smoothed
the linear power spectrum with a Gaussian of $1\,h^{-1}$Mpc.
Again $1500^3$ particles in a $1.5\,h^{-1}$Gpc box were employed and for each
particle, the value of the initial density, evaluated on the $1500^3$ grid,
was stored.
The particles were integrated to $z\simeq 0.55$ either using a particle-mesh
(PM) code (with a $1500^3$ mesh) or the Zel'dovich approximation.
Particles were then selected if their density field in the initial conditions
(extrapolated to $z=0$ using linear theory) exceeded some threshold.
In this manner the simulations mimic the analytic calculation closely.

In the analytic calculation we also used a linear theory power spectrum
smoothed with a $1\,h^{-1}$Mpc Gaussian, and we set $b_n$ assuming
$F(\delta)\propto \Theta(\delta-\delta_c)$, i.e.
\begin{eqnarray}
  b_1 &=& \sqrt{\frac{2}{\pi}}\left[ \sigma
  \  {\rm erfc}\left(\frac{\delta_c}{\sqrt{2}\,\sigma}\right)\right]^{-1}
    e^{-\delta_c^2/2\sigma^2} \to \frac{\delta_c}{\sigma^2} \\
  b_2 &=& \sqrt{\frac{2}{\pi}}\left[ \frac{\sigma^3}{\delta_c}
  \  {\rm erfc}\left(\frac{\delta_c}{\sqrt{2}\,\sigma}\right)\right]^{-1}
    e^{-\delta_c^2/2\sigma^2} \to \frac{\delta_c^2}{\sigma^4}
\end{eqnarray}
\citep{Sza88,Mat11} where the limits shown are for $\delta_c\gg 1$ and can be
compared to the leading order behavior in
Eqs.~(\ref{eqn:blag1}, \ref{eqn:blag2})
with $\nu=\delta_c/\sigma$.  We have chosen $\delta_c$ such that the
large-scale bias is approximately $1.6$,
as for the halo sample in Fig.~\ref{fig:cmp_zel}.

\begin{figure}
\begin{center}
\resizebox{3.3in}{!}{\includegraphics{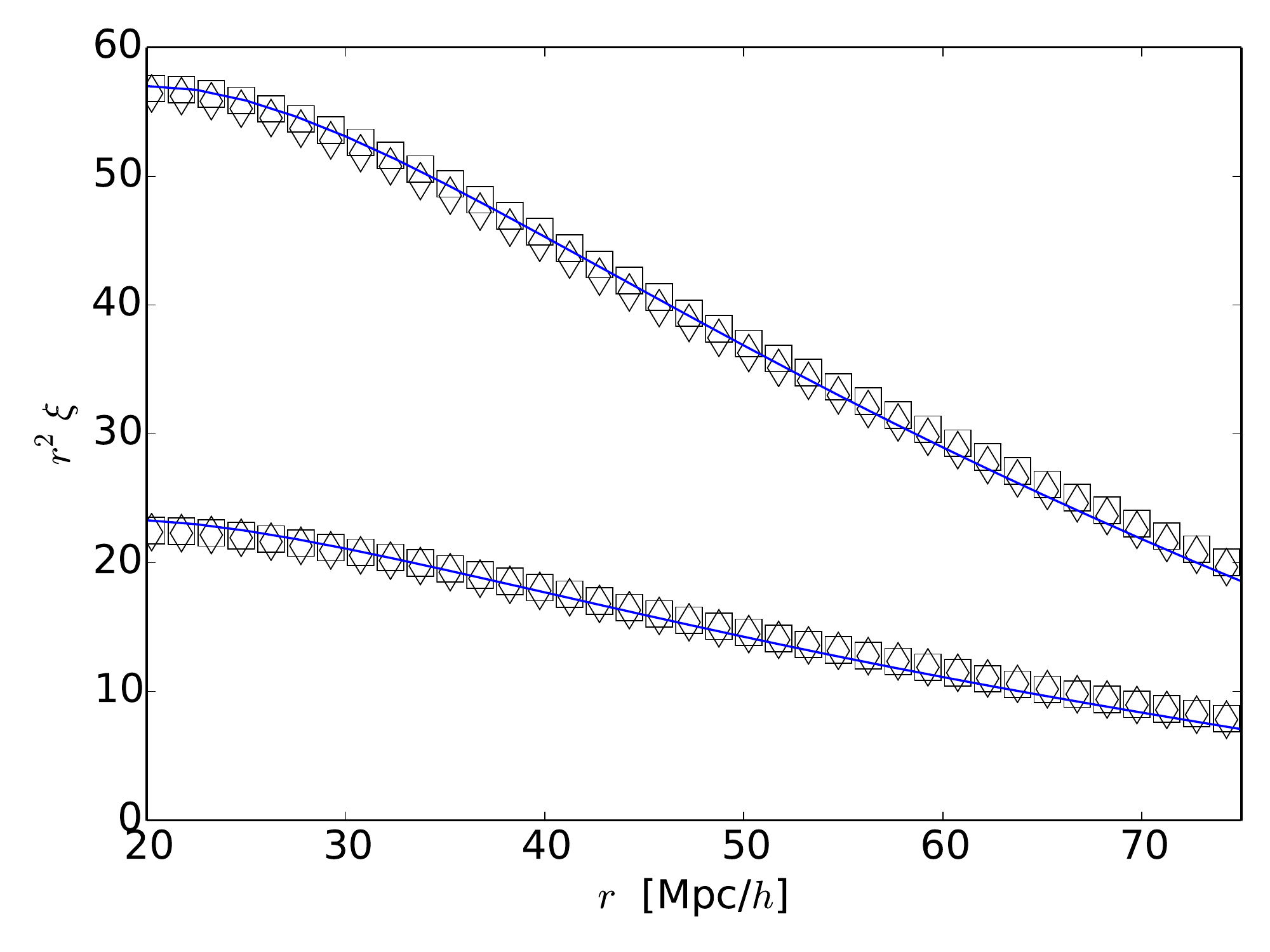}}
\resizebox{3.3in}{!}{\includegraphics{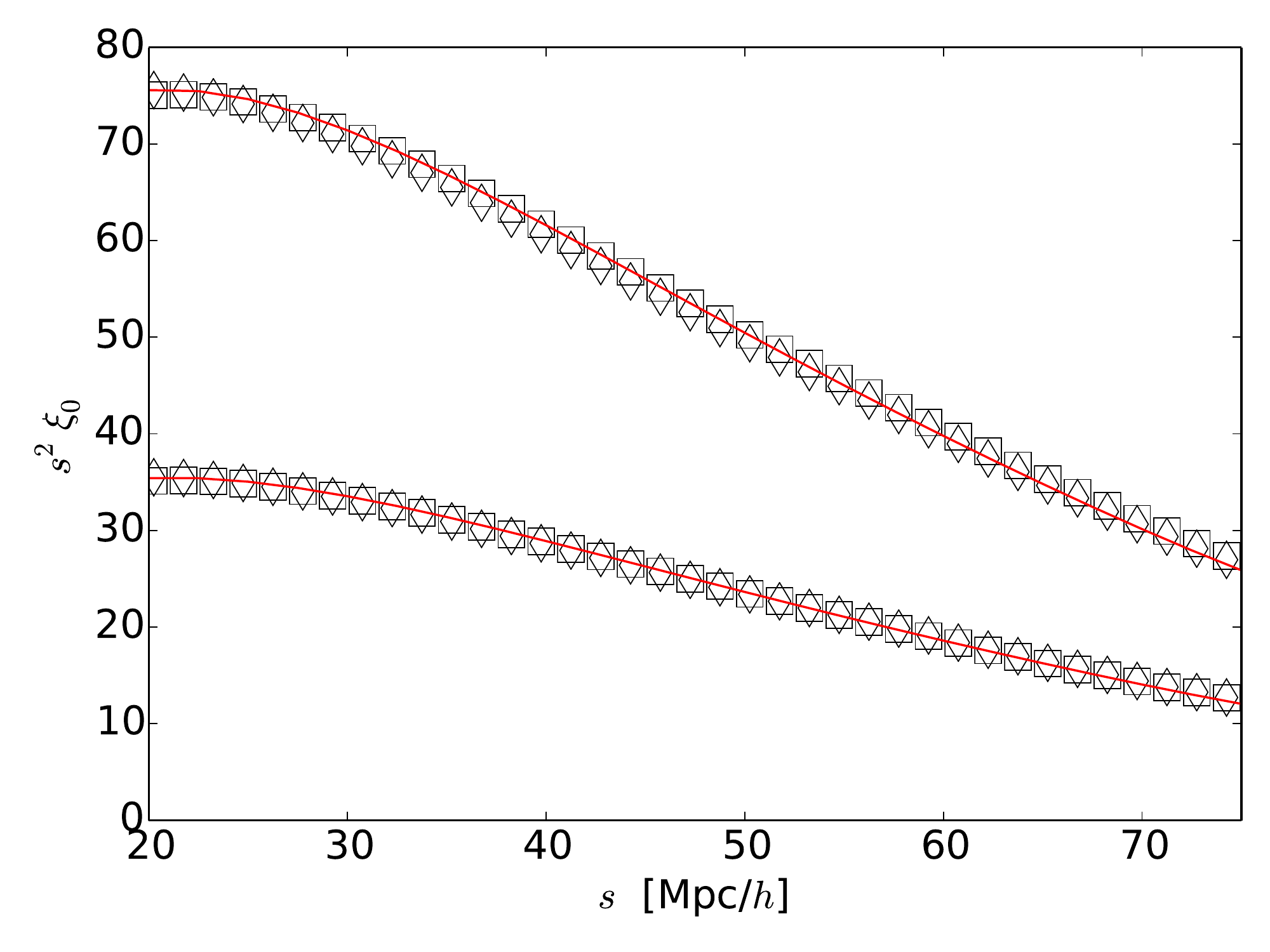}}
\resizebox{3.3in}{!}{\includegraphics{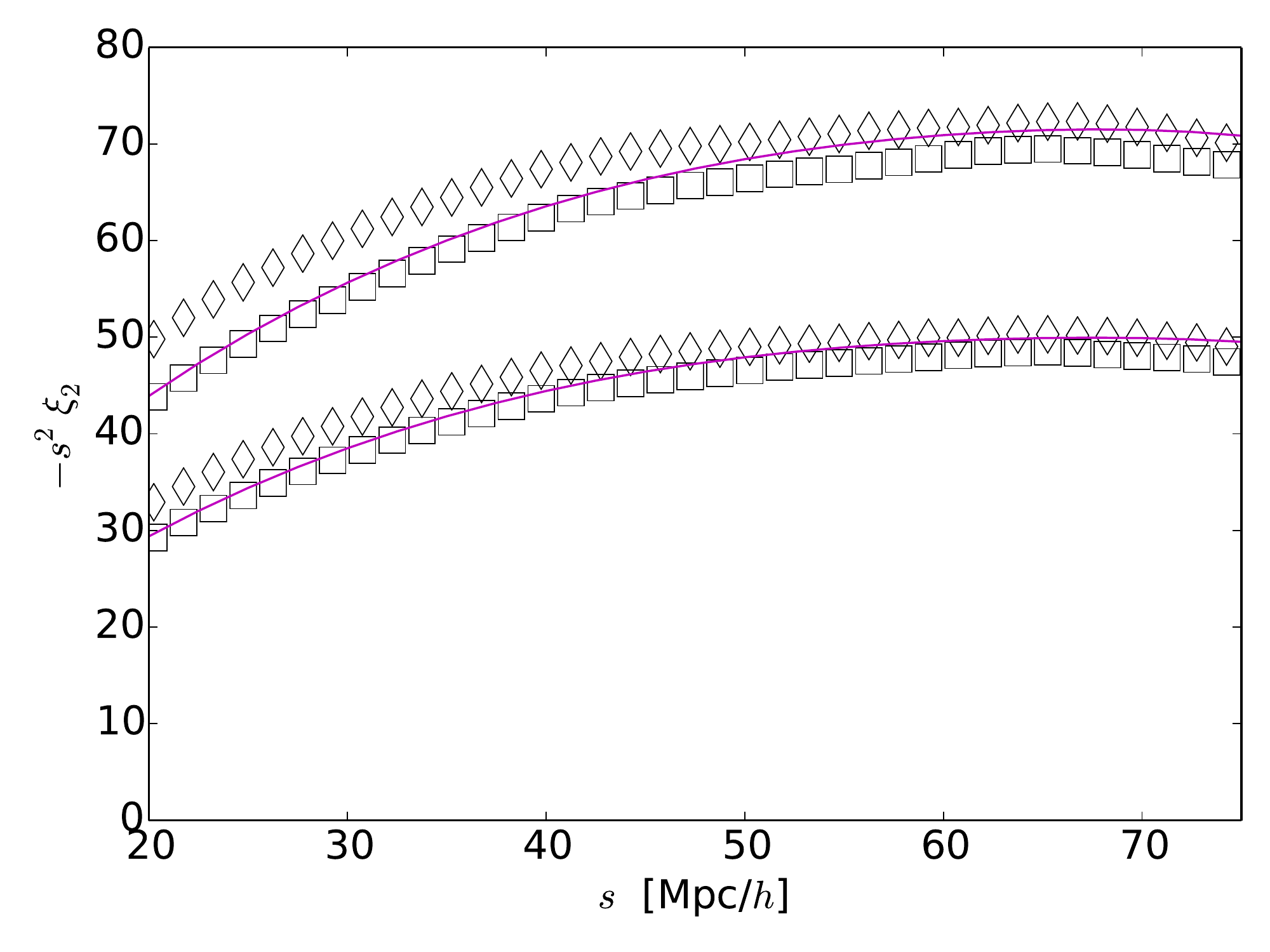}}
\end{center}
\caption{The real-space correlation function (top) and the monopole (middle)
and quadrupole (bottom) moments of the redshift-space correlation function
for our simulations based on thresholding.  In each panel the lines represent
the analytic model, the squares the results of the Zel'dovich simulations and
the diamonds the results of the PM simulations.  The upper curves/points are
for points with $\delta_{\rm lin}$ above a threshold while the lower
curves/points are for all $\delta_{\rm lin}$.  There are no free parameters
in this comparison!  Note that the quadrupole moment is more sensitive to the
full non-linear evolution than either the monopole or real-space correlation
function.}
\label{fig:pmz}
\end{figure}

Fig.~\ref{fig:pmz} shows the real-space correlation function and the monopole
and quadrupole moments of the redshift-space correlation function for all
three methods, focusing on intermediate scales.
The lower curves/points show the results for the matter field while the upper
curves/points show the results for all particles with initial $\delta$ above
a threshold.
Note that there are no free parameters in this comparison!
The agreement between the N-body results, the Zel'dovich simulations and the
theory is excellent for the real-space correlation function and the monopole
of the redshift-space correlation function.
The agreement remains good (though not perfect) for the quadrupole of the
matter field, but is less good for particles selected by initial density.
In particular the PM results show the same qualitative difference from the
Zel'dovich results as was found in the TreePM runs with halos.
This suggests that the mismatch in the halo quadrupole that we are seeing
in Figs.~\ref{fig:cmp_zel} and \ref{fig:three_panel} is at least partly due
to inadequacies in the Zel'dovich prediction for the inter-halo relative
velocities \citep[see also][]{SetYok98,TasZal12b,TasZal12c}, though some
may be due to our assumption of local Lagrangian bias.
We discuss non-local bias next.

\section{Beyond local bias}
\label{sec:nonlocal}

The failure of our model to match the quadrupole moment on small and
intermediate scales may be due in part to our assumption of local Lagrangian
bias.  While this approximation has received support from N-body simulations
\citep{RotPor11,Bal12,Chan12,WanSza12}
it must break down at some level.

Perhaps the simplest modification to our formalism would be to allow a
$q$-dependence to the bias coefficients, $b_n$.  For example we could
consider $b_1\to b_1[1+q_\star^2/q^2]$.  For suitably chosen $q_\star$,
such a modification can improve the agreement of the quadrupole at
intermediate scales ($s\approx 50\,h^{-1}$Mpc) but it changes the
monopole in a manner qualitatively similar to the overshoot seen in iPT
in Fig.~\ref{fig:cmp_zel}.
In general a large enough modification to create agreement for the
quadrupole removes the good agreement with the monopole.
However it is possible to adjust $q_\star$ so that the disagreement for
the monopole is on small scales ($s<30\,h^{-1}$Mpc) while the agreement
for the quadrupole is improved non-negligbly over the range
$30<s<75\,h^{-1}$Mpc.  Indeed, by ``softening'' the bias using a form like
$b_1\to b_1[1+q_\star^2/(q^2+\varepsilon^2)]$ both the monopole and
quadrupole can be made to agree with the N-body results to better than 5
per cent for $s>30\,h^{-1}$Mpc, though the theory rapidly departs from the
N-body results for smaller scales.

We also expect that terms involving e.g.~the tidal tensor, can become
important for high mass halos \citep{SheChaSco12}. 
Such terms are naturally quadrupolar in nature and may affect the predictions.
While shear terms are naturally induced by gravitational evolution, here we
are interested in any dependence in the initial conditions.

Suppose we extend $F(\delta)$ to also include a tidal-shear dependence,
$F(\delta,s^2)$, as discussed for example by \citet{McDRoy09}?
We can Fourier transform on $s^2$ as well, generating a term
$\exp[i\zeta s^2]$.  Since $s^2=s_{ij}s_{ij}$, with
\begin{equation}
  s_{ij}(\mathbf{k}) = \left(\frac{k_ik_j}{k^2}-\frac{1}{3}\delta_{ij}\right)
  \delta(\mathbf{k}) 
\end{equation}
is already quadratic in $\delta$, in the exponential it appears\footnote{When
using the cumulant theorem, bear in mind that $s^2$ is not Gaussian.} only
multiplied by other expectation values.
Throughout we shall subtract the mean,
$\langle s^2\rangle=(2/3)\langle\delta^2\rangle$, from $s^2$.
Working to lowest order in $\zeta$ the additional terms to be added inside
the $[\cdots]$ in Eq.~(\ref{eqn:K_expanded}) go as
\begin{equation}
 \left\langle s_1^2s_2^2    \right\rangle \quad , \quad
 \left\langle s^2\delta^2   \right\rangle \quad , \quad
 \left\langle s^2\delta\Psi \right\rangle \quad , \quad
 \left\langle s^2\vPsi^2    \right\rangle
\end{equation}
since the other terms involve the expectation value of an odd number of
Gaussian fields.
The relevant formulae can be found in the appendix.  The terms are
shear-density correlations:
$-i[\zeta_1\lambda_2^2+\zeta_2\lambda_1^2]
 \left\langle s_{ij}(\vq_1)\delta(\vq_2)\right\rangle^2$;
shear-shear terms: $-(\zeta_1^2+\zeta_2^2)\langle s^2\rangle$ and
$-\zeta_1\zeta_2\langle s_1s_2\rangle^2$;
shear-displacement terms:
$(\zeta_1+\zeta_2)\langle s_{ij}(\vq_1)k_m\Psi_m(\vq_2)\rangle^2$
and cross-terms
$-2i(\zeta_1\lambda_2+\zeta_2\lambda_1)
 \langle s_{ij}(\vq_1)\delta(\vq_2)\rangle
 \langle s_{ij}(\vq_1)k_m\Psi_m(\vq_2)\rangle$.
These last terms include contributions with $\hk\cdot\hq$ and
$(\hk\cdot\hq)^2$ into Eq.~(\ref{eqn:L_ZA}) and appear to be the best bet
for influencing the quadrupole.

The terms involving $s^2$ are very small, as expected since they are higher
order in $P_L$.  Even allowing for arbitrary prefactors in front of the terms,
the halo quadrupole cannot be matched without spoiling the agreement with the
redshift-space monopole and real-space correlation function.
This is because the terms which enter contribute approximately the same
amount to the monopole as to the quadrupole (as was the case with the
non-shear terms).
It thus appears that the lack of shear terms in the bias function is not the
reason for the discrepancy seen in Fig.~\ref{fig:cmp_zel}.

One may take a more general approach.  Within the context of the
Zel'dovich approximation, the terms appearing in the square brackets
in Eq.~(\ref{eqn:L_ZA}) will be functions of $q$ and will be contracted
with various factors of $\vk$.  There will be scalars, like $\xi_R$,
vectors, like $\mathbf{U}$, and tensors of various ranks.  The vectors
must be proportional to $\hq_i$.  The rank-2 tensors must go as a sum of
terms like $\delta_{ij}$ and $\hq_i\hq_j$, and similarly for higher rank
objects.  The most general biasing scheme would therefore consist of all
such terms, with general dependence on $q$.
We have not undertaken an exploration of this large parameter space, but
our experience above suggests that any such terms will contribute
approximately equally to the monopole and the quadrupole moment, making it
difficult to substantially adjust the quadrupole on small scales without
spoiling the agreement seen for the monopole.  As we saw above when we
modified $b_1\to b_1(1+q_\star^2/q^2)$, it is possible to improve the
level of agreement in some cases, though we do not posses a theory which
predicts the required functional form at present.

On the basis of these investigations it appears that the disagreement
between the Zel'dovich prediction for the quadrupole moment of the
redshift-space, halo correlation function and that measured in N-body
simulations may be due to simplifications inherent in the Zel'dovich
approximation itself.  Perhaps the relative velocities predicted from the
lowest order displacement field are not as accurate for larger $\delta_L$.
Consistent with this view, we note that CLPT (which goes to next order in
Lagrangian perturbation theory) does perform (very slightly) better than
the Zel'dovich approximation in the range $40<s<70\,h^{-1}$Mpc.
Higher order LPT calculations are also known to give more accurate values
for the moments \citep[e.g.][]{MunSahSta94}.
However, the convergence appears to be very slow at best.

\section{Zel'dovich streaming model (ZSM)}
\label{sec:zsm}

If the pure Zel'dovich calculation cannot match the small-scale quadrupole
moment of the halo correlation function, can it be part of an extended model
which can?  The tests above suggest that the  failure of the Zel'dovich
approximation lies in the pairwise velocity distribution predicted by the
model.   \citet{ReiWhi11} showed that the halo pairwise velocity
distribution was quite well approximated by a Gaussian.  What if we enforce
this functional form (the ``Gaussian streaming model''), using the Zel'dovich
approximation to compute the ingredients: the real-space clustering of
biased tracers, the mean infall velocity and the velocity dispersion?
Specifically we assume
\begin{equation}
  1 + \xi^s(s_\perp, s_\parallel) =
  \int \dfrac{dy}{[2\pi]^{1/2}\sigma_{12}}
  [1 + \xi(r)]
  \exp\left\{-\frac{[s_\parallel-y-\mu v_{12}]^2}{2\sigma_{12}^2}\right\}\ ,
\label{eq:streaming-xi-s}
\end{equation}
with $\xi(r)$, $v_{12}$ and $\sigma_{12}$ from the analytic theory.
This expression simply enforces pair counting, assuming that the functional
form of the velocity distribution is Gaussian, centered at $\mu v_{12}$;
the mean LOS velocity between a pair of tracers as a function of their real
space separation.
We have just shown that the Zel'dovich approximation works well for the
real-space correlation function of biased tracers, such as halos.
The scale-dependence of the velocity dispersions\footnote{If our goal is to
model the clustering of galaxies, then we must include a phenomenological
model for the finger-of-god effect.  \citet{Rei12} showed that a single extra
parameter -- an isotropic velocity dispersion -- sufficed to model
fingers-of-god on large scales.}
is well predicted by linear theory \citep{ReiWhi11,WanReiWhi13}.
Thus the only missing ingredient is the mean infall velocity.

We can use the method described in \citet{WanReiWhi13} to compute $v_{12}$
within the Zel'dovich approximation.  One simply adds a term
$\mathbf{J}\cdot\mathbf{\dot{\Delta}}$ to the exponent in Eq.~(\ref{eqn:K})
and computes derivatives with respect to $\mathbf{J}$,  This is a subset of
the calculation presented in \citet{WanReiWhi13}:
\begin{equation}
  \mathbf{v}_{12} = \left[1+\xi\right]^{-1}\int d^3q
  \ \mathbf{M}(\mathbf{r},\mathbf{q})
\end{equation}
with\footnote{There is a typographical error in the subscripts in Eqs.~(31, 32)
of \citet{WanReiWhi13} that is corrected here.  The combination
$k_ik_jU_i\dot{A}_{in}$ should have been $k_ik_jU_i\dot{A}_{jn}$ and
$G_{ij}U_i\dot{A}_{in}$ should have been $G_{ij}U_i\dot{A}_{jn}$.  None of
the results in \citet{WanReiWhi13} were affected.}
\begin{eqnarray}
  M_n &=& \frac{f}{(2\pi)^{3/2}|A|^{1/2}}
    e^{-(1/2)(q_i-r_i)A^{-1}_{ij}(q_j-r_j)} \nonumber \\
    &\times& \bigg\{ - g_i A_{in}
    + 2b_1\left[  U_n - U_iG_{ij}A_{jn} \right]
    - 2b_2 g_i U_i U_n \nonumber \\
    &-& b_1^2\left[2 g_i U_i U_n + \xi_L g_i A_{in} \right]
     + 2b_1 b_2 \xi_L U_n \bigg\} \ ,
\end{eqnarray}
Note that we are interested in the line-of-sight velocity, so we
require only the component of $\mathbf{M}$ along $\mathbf{r}$.

\begin{figure}
\begin{center}
\resizebox{3.3in}{!}{\includegraphics{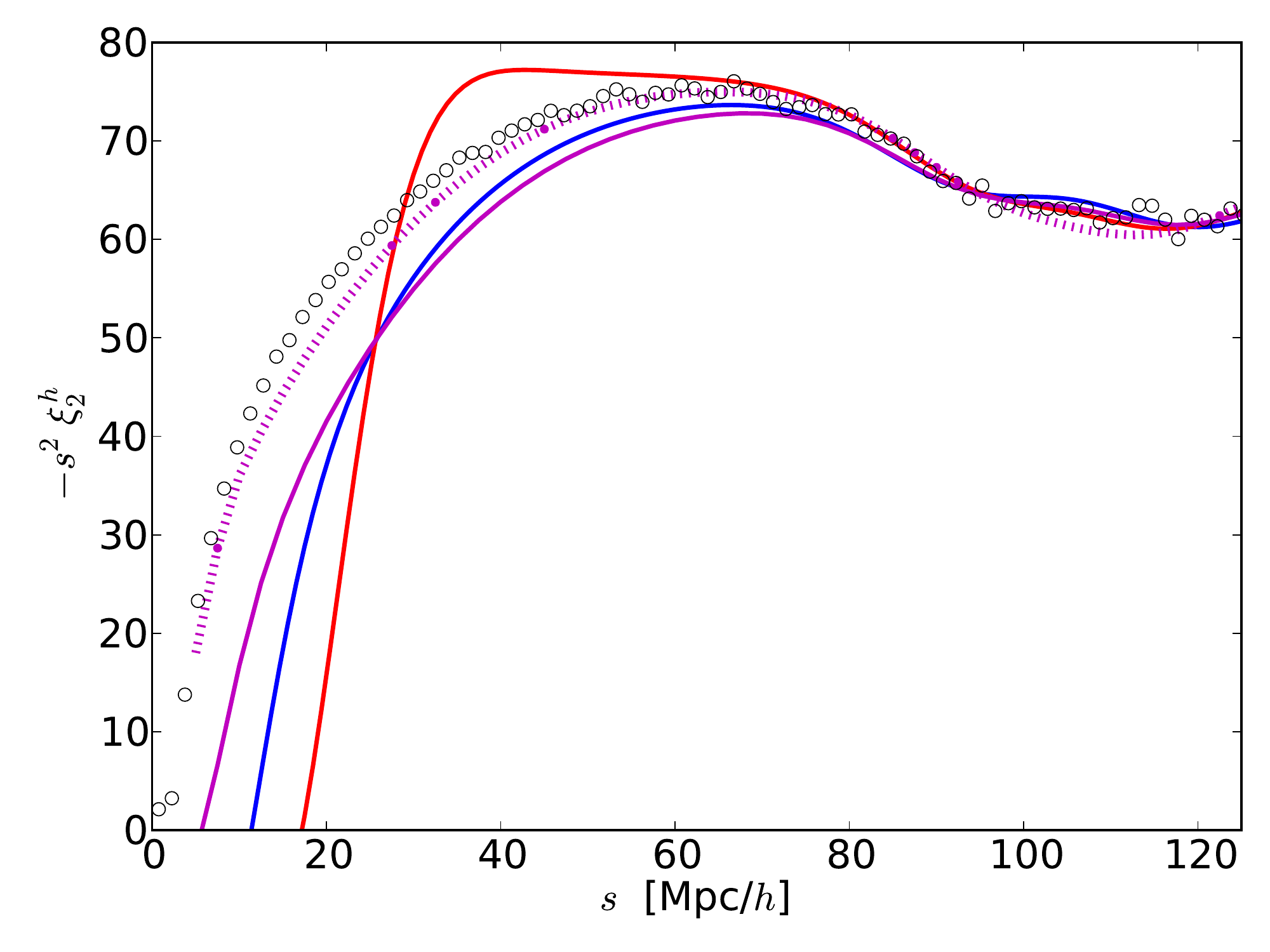}}
\end{center}
\caption{The quadrupole moment of the halo correlation function,
multiplied by $s^2$.  This is the same as the lower right panel
of Fig.~\ref{fig:cmp_zel}.}
\label{fig:quadrupole}
\end{figure}

The predictions of the Zel'dovich streaming model (ZSM) are shown in
Fig.~\ref{fig:cmp_zel} as the dotted magenta lines.
They are indistinguishable from the Zel'dovich model above except for
the lower right panel (halo quadrupole).  We reproduce the quadrupole
results in Fig.~\ref{fig:quadrupole} where we see the model improves
upon the Zel'dovich calculation significantly at small separation, even
though it only involves simple, one-dimensional integrals of the linear
theory power spectrum.
Above $15\,h^{-1}$Mpc the ZSM prediction for the quadrupole is within
10 per cent of the N-body result.  Above $30\,h^{-1}$Mpc it is within
5 per cent of the N-body result and above $50\,h^{-1}$Mpc it is within
1 per cent of the N-body result.  For the halo sample shown in
Fig.~\ref{fig:three_panel} the corresponding numbers are 4 per cent,
1 per cent and 1 per cent.

\section{Discussion}
\label{sec:discussion}

The Zel'dovich approximation \citep{Zel70} remains one of our most
powerful analytic models of large-scale structure.  We have presented
a derivation of correlation function, in real- and redshift-space, within
the Zel'dovich approximation including an analytic inversion of the
Lagrangian correlator which appears as the fundamental ingredient of the
model.  The resulting integral expression is exact within the context of
the Zel'dovich approximation and can be rapidly evaluated using quadratures.
We have compared the Zel'dovich calculation to higher-order Lagrangian
schemes and shown that the perturbation theory appears to be converging quickly
\citep[see also][]{Tas14a}.  All of the Lagrangian perturbation theories
provide a good match to the N-body results on scales above $20\,h^{-1}$Mpc.

The calculation has been extended to include biased tracers of the density
field, including terms which are third order in the linear theory power
spectrum.  We have also considered non-local bias terms such as a dependence
on tidal shear in the initial field.  We find that these higher order terms
are generally very small.  The Zel'dovich approximation provides a very
good fit to the real-space correlation function of halos found in N-body
simulations, and to the monopole of the redshift-space correlation function.
However, it does not match the quadrupole moment below $75\,h^{-1}$Mpc.
Modifications to the bias terms which introduce a scale-dependence can
improve the agreement with the N-body results, but usually at the expense of
worsening the agreement in the monopole.
We have argued that at least some of this disagreement is an issue with the
approximation itself, and that it predicts the wrong pairwise velocity
distribution for biased tracers.

Finally we have used the Zel'dovich approximation to compute the ingredients
of the Gaussian streaming model of \citet{ReiWhi11}.  We find that this
hybrid model, which we refer to as the Zel'dovich streaming model and which
involves only simple integrals of the linear theory power spectrum, provides
a good match to the N-body measurements down to tens of Mpc.

\vspace{0.2in}
M.W. would like to thank Matt McQuinn for numerous helpful conversations
about this work and the referee, Adrian Melott, for comments which improved
the draft.  M.W. is supported by NASA.
This work made extensive use of the NASA Astrophysics Data System and
of the {\tt astro-ph} preprint archive at {\tt arXiv.org}.
The analysis made use of the computing resources of the National Energy
Research Scientific Computing Center.


\newcommand{\aj}{AJ}
\newcommand{\apj}{ApJ}
\newcommand{\apjs}{ApJ Suppl.}
\newcommand{\mnras}{MNRAS}
\newcommand{\araa}{ARA{\&}A}
\newcommand{\aap}{A{\&}A}
\newcommand{\pre}{PRE}
\newcommand{\prd}{Phys. Rev. D}
\newcommand{\apjl}{ApJL}
\newcommand{\physrep}{Physics Reports}
\newcommand{\nat}{Nature}

\setlength{\bibhang}{2.0em}
\setlength\labelwidth{0.0em}
\bibliography{biblio}

\begin{appendix}

\section{The shear terms}
\label{sec:shear}

In order to compute the contributions from any $s^2$ terms to the
correlation function(s) we need to evaluate several two-point functions.
The simplest expectation value is
$\langle s^2(\vq)\rangle=(2/3)\langle\delta^2\rangle$, independent of
position.  The next simplest is
\begin{equation}
  \left\langle s_{ij}(\vq_1)\delta(\vq_2)\right\rangle =
  \left(\frac{1}{3}\delta_{ij}-\hq_i\hq_j\right) \mJ_1(q)
\end{equation}
where $\vq=\vq_2-\vq_1$ and $\mJ_1$ is defined below.  Then, for example,
$\left\langle s_{ij}(\vq_1)\delta(\vq_2)\right\rangle
 \left\langle s_{ij}(\vq_1)\delta(\vq_2)\right\rangle=(2/3)\mJ_1^2(q)$.
By similar logic
\begin{equation}
  \left\langle s_{ij}(\vq_1)\Psi_m(\vq_2)\right\rangle =
  \delta_{ij}\hq_m\mJ_2+\left[\delta_{jm}\hq_i+\delta_{im}\hq_j\right]\mJ_3
  + \hq_i\hq_j\hq_m \mJ_4
\end{equation}
with
\begin{eqnarray}
  && \left\langle s_{ij}(\vq_1)\Psi_m(\vq_2)\right\rangle
  \left\langle s_{ij}(\vq_1)\Psi_n(\vq_2)\right\rangle =
  2\mJ_3^2\delta_{mn}+\hq_m\hq_n \times \nonumber \\
  &&\left[ 2\mJ_2^2+2\mJ_3^2+\mJ_4^2 + 3\mJ_2\mJ_3+2\mJ_2\mJ_4+4\mJ_3\mJ_4
    \right]
\end{eqnarray}
and
\begin{equation}
  \left\langle s_{ij}(\vq_1)\delta(\vq_2)\right\rangle
  \left\langle s_{ij}(\vq_1)\Psi_m(\vq_2)\right\rangle =
  -\frac{2}{3}\hq_m\ \mJ_1\left[2\mJ_3+\mJ_4\right] \quad .
\end{equation}
Also
\begin{eqnarray}
  \left\langle s_{ij}(\vq_1)s_{mn}(\vq_2)\right\rangle &=&
  \delta_{ij}\delta_{mn}\mJ_5 +
  \left[\delta_{im}\delta_{jn}+\delta_{in}\delta_{jm}\right]\mJ_6 \nonumber \\
  &+& \left[\delta_{ij}\hq_m\hq_n+\delta_{mn}\hq_i\hq_j\right]\mJ_7\nonumber \\
  &+& \left[\delta_{im}\hq_j\hq_n+\delta_{in}\hq_j\hq_m+
            \delta_{jm}\hq_i\hq_n+\delta_{jn}\hq_i\hq_m \right]\mJ_8\nonumber\\
  &+& \hq_i\hq_j\hq_m\hq_n\mJ_9
\end{eqnarray}
and thus the contraction
\begin{eqnarray}
 && \langle s_{ij}(\vq_1)s_{mn}(\vq_2)\rangle
 \langle s_{ij}(\vq_1)s_{mn}(\vq_2)\rangle = \nonumber \\
 && 9\mJ_5^2 + 24\mJ_6^2 + 8\mJ_7^2 + 24\mJ_8^2 + \mJ_9^2 \nonumber \\
 &+& 12\mJ_5\mJ_6  + 12\mJ_5\mJ_7 + 8\mJ_5\mJ_8 + 2\mJ_5\mJ_9 \nonumber \\
 &+& 8\mJ_6\mJ_7 + 32\mJ_6\mJ_8 + 4\mJ_6\mJ_9 \nonumber \\
 &+& 16\mJ_7\mJ_8 + 2\mJ_7\mJ_9 + 8\mJ_8\mJ_9
\end{eqnarray}
The terms $\mJ_i$ all have argument $q$ and are simple integrals over the
linear theory power spectrum,
\begin{eqnarray}
  \mJ_1(q) &=& \int\frac{k^2\,dk}{2\pi^2}P_L(k)\ j_2(kq) \\
  \mJ_2(q) &=& \int\frac{k\,dk}{2\pi^2}P_L(k)
  \ \left[\frac{2}{15}j_1(kq)-\frac{1}{5}j_3(kq)\right] \\
  \mJ_3(q) &=& \int\frac{k\,dk}{2\pi^2}P_L(k)
  \ \left[-\frac{1}{5}j_1(kq)-\frac{1}{5}j_3(kq)\right] \\
  \mJ_4(q) &=& \int\frac{k\,dk}{2\pi^2}P_L(k)\ j_3(kq) \\
  \mJ_5(q)&=&\int\frac{k^2\,dk}{2\pi^2}\ P_L(k)\frac{-14j_0-25j_2+24j_4}{315} \\
  \mJ_6(q)&=&\int\frac{k^2\,dk}{2\pi^2}\ P_L(k)\frac{  7j_0+ 5j_2- 2j_4}{105} \\
  \mJ_7(q)&=&\int\frac{k^2\,dk}{2\pi^2}\ P_L(k)\frac{        3j_2- 4j_4}{ 21} \\
  \mJ_8(q)&=&\int\frac{k^2\,dk}{2\pi^2}\ P_L(k)\frac{       -2j_2- 2j_4}{ 21} \\
  \mJ_9(q)&=&\int\frac{k^2\,dk}{2\pi^2}\ P_L(k)\frac{        -j_2+20j_4}{ 21}
\end{eqnarray}
where we have suppressed the $kq$ argument of the spherical Bessel functions
in the last few equations.  Note that $\mJ_5$ and $\mJ_6$ have non-zero limits
as $q\to 0$ but all of the other terms vanish in this limit.
It is easy to show that
$\mJ_5\to -(2/45)\langle\delta^2\rangle$ and
$\mJ_6\to (1/15)\langle\delta^2\rangle$ so that
\begin{equation}
  \frac{\left\langle s_{ij}(0)s_{mn}(0)\right\rangle}{\langle\delta^2\rangle}
  \to -\frac{2}{45}\delta_{ij}\delta_{mn} + \frac{1}{15}
  \left[\delta_{im}\delta_{jn}+\delta_{in}\delta_{jm}\right]
\end{equation}
as $q\to 0$ and therefore
\begin{equation}
  \left\langle s_{ij}(0)s_{ij}(0)\right\rangle \to
    \frac{2}{3} \langle\delta^2\rangle
\end{equation}
which agrees with our earlier result.

Numerically the largest contributions on scales above $20\,h^{-1}$Mpc
are from $\mJ_3$ and $\mJ_4$.  The next largest (in absolute magnitude) are
$\mJ_1$, $\mJ_2$ and $\mJ_9$ and then $\mJ_8$.  All terms are very smoothly
varying functions of $q$, like their counterparts $\sigma^2(q)$ and $U(q)$
shown in Fig.~\ref{fig:qf}.

\vspace{0.2in}
\noindent{\bf Erratum:}\newline
\vspace{0.1in}

The coefficients given in the integrals $\mJ_5-\mJ_9$ are incorrect.  They
should read
\begin{eqnarray}
  \mJ_5(q)&=&\int\frac{k^2\,dk}{2\pi^2}\ P_L(k)\frac{-14j_0-40j_2+ 9j_4}{315} \\
  \mJ_6(q)&=&\int\frac{k^2\,dk}{2\pi^2}\ P_L(k)\frac{  7j_0+10j_2+ 3j_4}{105} \\
  \mJ_7(q)&=&\int\frac{k^2\,dk}{2\pi^2}\ P_L(k)\frac{        4j_2- 3j_4}{ 21} \\
  \mJ_8(q)&=&\int\frac{k^2\,dk}{2\pi^2}\ P_L(k)\frac{       -3j_2- 3j_4}{ 21} \\
  \mJ_9(q)&=&\int\frac{k^2\,dk}{2\pi^2}\ P_L(k) j_4
\end{eqnarray}
The conclusions are unchanged.

\end{appendix}

\label{lastpage}
\end{document}